\documentclass[
    nofootinbib,
    reprint,
    aps,
    prapplied,
    amsmath,
    amssymb,
    superscriptaddress,
    longbibliography,
    english
]{revtex4-2}
\usepackage[normalem]{ulem}
\usepackage{graphicx}   
\usepackage{physics}
\usepackage{siunitx}
\usepackage{wasysym}
\usepackage{float}
\usepackage{color}
\usepackage{bbold}
\usepackage{amsmath}
\usepackage{amsfonts}
\usepackage{amssymb}
\usepackage[title]{appendix}
\usepackage[dvipsnames]{xcolor}
\usepackage[hidelinks,colorlinks=true,linkcolor=blue,filecolor=blue,urlcolor=blue,citecolor=blue]{hyperref}
\usepackage[all]{hypcap}

\usepackage{mathtools}
\usepackage{soul}
\usepackage{float}
\makeatletter
\let\newfloat\newfloat@ltx
\makeatother

\usepackage{algpseudocode}
\usepackage{algorithm}
\usepackage{blindtext}

\newif\ifdraft
\draftfalse

\def \param {\mathcal{K}}

\begin{document}

\title{Sensitivity-Adapted Closed-Loop Optimization for High-Fidelity Controlled-Z Gates in Superconducting Qubits}

\author{N. J. Glaser}
\email{niklas.glaser@wmi.badw.de}
\affiliation{Technical University of Munich, TUM School of Natural Sciences,
Department of Physics, Garching 85748, Germany}
\affiliation{Walther-Mei{\ss}ner-Institut, Bayerische Akademie der Wissenschaften, 85748 Garching, Germany}
\author{F. A. Roy}
\affiliation{Walther-Mei{\ss}ner-Institut, Bayerische Akademie der Wissenschaften, 85748 Garching, Germany}
\affiliation{Theoretical Physics, Saarland University, 66123 Saarbr\"ucken, Germany}
\author{I. Tsitsilin}
\author{L. Koch}
\author{N. Bruckmoser}
\author{J. Schirk}
\author{J. H. Romeiro}
\author{G. B. P. Huber}
\author{F. Wallner}
\author{M. Singh}
\author{G. Krylov}
\author{A. Marx}
\author{L. Södergren}
\author{C. M. F. Schneider}
\author{M. Werninghaus}
\affiliation{Technical University of Munich, TUM School of Natural Sciences,
Department of Physics, Garching 85748, Germany}
\affiliation{Walther-Mei{\ss}ner-Institut, Bayerische Akademie der Wissenschaften, 85748 Garching, Germany}

\author{S. Filipp}
\email{stefan.filipp@wmi.badw.de}
\affiliation{Technical University of Munich, TUM School of Natural Sciences,
Department of Physics, Garching 85748, Germany}
\affiliation{Walther-Mei{\ss}ner-Institut, Bayerische Akademie der Wissenschaften, 85748 Garching, Germany}
\affiliation{Munich Center for Quantum Science and Technology (MCQST), Schellingstra\ss e 4, 80799 München, Germany}

\begin{abstract}
    Achieving fast and high-fidelity qubit operations is crucial for unlocking the potential of quantum computers.
  In particular, reaching low gate errors in two-qubit gates has been a long-standing challenge in the field of superconducting qubits due to their typically long duration relative to coherence times. 
  To realize fast gates, we utilize the hybridization between fixed-frequency superconducting qubits with a strongly interacting coupler mode that is tunable in frequency.
    To reduce population leakage during required adiabatic passages through avoided level crossings, we employ a sensitivity-adaptive closed-loop optimization method to design complex pulse shapes.
    We compare the performance of Gaussian-square, Fourier-series, and piecewise-constant-slope (PiCoS) pulse parametrizations and are able to reach 99.9\% controlled-Z gate fidelity using a \SI{64}{\nano\second} long Fourier-series pulse defined by only seven parameters. 
   These high-fidelity values are achieved by analyzing the optimized pulse shapes to identify and systematically mitigate signal-line distortions in the experiment. To improve the convergence speed of the optimization we implement an adaptive cost function, which continuously maximizes the sensitivity. The demonstrated method can be used for tune-up and recalibration of superconducting quantum processors.
\end{abstract}
\maketitle
\maketitle

\section{Introduction}\label{sec:introduction}
In the rapidly developing field of quantum computing, superconducting qubits have emerged as a promising architecture for scalable quantum processors that operate with high speed~\cite{Kjaergaard2020, krinner2022realizing, google2023suppressing}. 
Superconducting qubits provide a complete gate set by using single qubit gate rotations of fixed angle, arbitrary Z-rotations, and an entangling gate between the individual qubits.
Despite significant improvements in coherence times of these qubits~\cite{place2021new, wang2022towards, gordon2022environmental, deng2023titanium, bal2024systematic}, the fidelity of two-qubit operations such as the controlled-Z (CZ) gates remains a limiting factor.

Many different schemes for realizing two-qubit gates on superconducting qubits have been demonstrated, using parametric coupler interactions~\cite{bialczak2011fast, Mckay2016, huber2024parametric}, cross-resonance drives~\cite{rigetti2010fully,chow2011simple}, microwave-activated controlled phases~\cite{chow2013microwave}, resonator induced phase gates (RIP)~\cite{paik2016experimental} and the utilization of higher qubit states~\cite{foxen2020demonstrating, Sung2021, Marxer2023}.
To implement CPHASE gates with the latter, the transition between the states $\ket{11}$, where both qubits are excited, and $\ket{20}$, where one qubit is in its second-excited state, is used by tuning the frequency of the qubits and potentially also the coupler.
In this scheme, high-fidelity gates with fidelities exceeding \SI{99.8}{\percent} have been demonstrated using analytically derived pulse shapes~\cite{Marxer2023}.

We investigate an alternative implementation using fixed-frequency qubits that are not affected by flux-noise~\cite{Koch2007} and, therefore, achieve long coherence times.

To realize fast gates, we utilize the adiabatic hybridization between fixed-frequency superconducting qubits with a strongly interacting coupler mode~\cite{Yan2018, Collodo2020,  Xu2020, Stehlik2021,glaser2023controlled, Rui2023Realization}.
By varying the frequency of the coupler mode, a ZZ-type interaction between the qubits is activated. This interaction is quantified by the conditional frequency shift $\xi=(\tilde{\omega}_{11} - \tilde{\omega}_{01} - \tilde{\omega}_{10} + \tilde{\omega}_{00})$ of the double-excited state, $\ket{11} \rightarrow e^{i \xi t}\ket{11}$, where $\tilde{\omega}_{\text{n}_1 \text{n}_2}$ denote the eigenfrequencies of the basis states $\ket{n_1n_2}$. To realize a high on-off ratio, $\xi$ must be small during idling operation and as large as possible during gate operation. The requirement for small $\xi$ during idling can be satisfied by exploiting the cancellation of two coupling paths between the qubits: one indirect via the tunable coupler and one through direct capacitive coupling. Due to their opposite signs, a zero $\xi$ configuration can be found with a suitable tunable-coupler frequency in the idling state~\cite{mundada2019suppression,Sete2021, Stehlik2021}.

The second condition is realized by the hybridization of higher excited qubit and coupler states.
The adiabatic tuning of the coupler frequency then allows to adjust the conditional frequency shift  $\xi$ over several orders of magnitude~\cite{Collodo2020, Xu2020,Chu2021, Stehlik2021}.
For a high-fidelity operation, the coupler trajectory must be shaped such that the correct amount of phase is accumulated and any state transfers caused by energy level hybridizations are reversed. 
One option is to use analytically derived pulse shape descriptions for adiabatic CZ gates~\cite{Martinis2014, Sung2021, Stehlik2021}. These, however, typically consider only a single avoided energy-level crossing to derive the pulse description. 

Identifying general analytical formulas suitable for the more complex energy structure that arises due to the higher qubit state interactions remains challenging, and one typically resorts to numerical open-loop optimization to efficiently compute pulse shapes~\cite{khaneja2005, reich2012monotonically, doria2011optimal}. 
In the transfer to experiments, mismatches between models and hardware, such as transfer functions
and uncertainties in model parameters may, however, lead to systematic deviations from the targeted system dynamics~\cite{Machnes2018, Wittler2021, werninghaus2021leakage}.
While parameters of the system model and of transfer functions can, in principle, all be calibrated individually~\cite{sheldon2016characterizing}, the large number of potentially correlated parameters leads to a complex optimization problem with low convergence.  
To overcome these challenges we instead perform a closed-loop pulse optimization of the CZ gate. 
 Based on a prior pulse shape with low complexity, we optimize pulse shapes with an extended number of parameters on the device, using a global cost function based on randomized benchmarking.  
 This allows us to find optimal parameters even in the presence of strong correlations between parameters. 
The optimization is able to converge to high-fidelity pulse shapes, correctly considering all avoided crossings and intrinsically correcting for transfer functions of the control lines.

\section{Coupler activated \\ conditional phase}

\begin{figure}[t]
    \centering
    \includegraphics{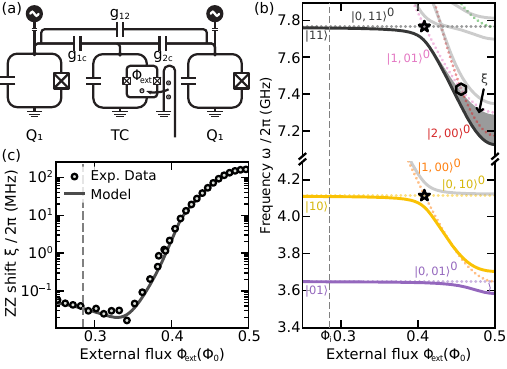}
    \caption{
        (a) Lumped-element circuit diagram of the system with two qubits ($Q_1$, $Q_2$) and a frequency-tunable coupler ($TC$). External drives are applied to each qubit for single-qubit control. The coupler is controlled via a current controllled flux line generating an external flux $\Phi_\text{ext}$ through the SQUID loop. The three elements are capacitively coupled with coupling strength $g_{ij}$. 
        (b) Energy-level diagram of the two-qubit system in the one-excitation and two-excitation manifolds.
        The solid lines show the adiabatic computational states $\ket{n_1n_2}$ as a function of the external flux $\Phi_\text{ext}$.
        The dotted lines show the system's bare states $\ket{n_c, n_1 n_2}^0$.
        The gray shaded area indicates the flux-dependent frequency shift $\xi$.
        The dashed line indicates the idle position of the coupler.
        (c) Frequency shift $\xi$ measured as a function of the extenal flux $\Phi_\text{ext}$
    }
    \label{fig:system_chip}
\end{figure}

The controlled-Z (CZ) gate is realized in a unit cell of a superconducting qubit architecture with a pair of fixed frequency transmon-type qubits ($Q_1$, $Q_2$) coupled via a flux-tunable coupling element $C$.
The corresponding circuit diagram shown in Fig.~\ref{fig:system_chip}(a) is described by the Hamiltonian
\begin{equation}
    \begin{aligned}                                                                                                                                                                                                                                                                           \\
        \hat{H} / \hbar              = & \sum_{i=1,2,c}\left(\omega_i \hat{a}_i^{\dagger} \hat{a}_i+\frac{\alpha_i}{2} \hat{a}_i^{\dagger} \hat{a}_i^{\dagger} \hat{a}_i \hat{a}_i\right) 
        \\
                          & -\sum_{i=1,2} g_{i c}\left(\hat{a}_i^{\dagger}-\hat{a}_i\right)\left(\hat{a}_c^{\dagger}-\hat{a}_c\right)                                        \\
                          & -g_{12}\left(\hat{a}_1^{\dagger}-\hat{a}_1\right)\left(\hat{a}_2^{\dagger}-\hat{a}_2\right) \,,
        \label{eq:system_ham}
    \end{aligned}
\end{equation}
with creation and annihilation operators $\hat{a}_i^{\dagger}$ and $\hat{a}_i$, frequencies $\omega_i$, and anharmonicities $\alpha_i$, with $i\in\{1,2\}$ for the qubits and $i=c$ for the tunable coupler. 
The frequency $\omega_c\left(\Phi_\text{ext}\right)$ of the coupler is tuned by threading an external magnetic flux $\Phi_{\text{ext}}$ through the SQUID loop of the coupler \cite{Krantz2019}.
The setup is described in Appendix~\ref{app:exp_setup} with the experimentally realized parameters summarized in Table~\ref{tab:qubit_paramters}.
 The elements are capacitively coupled with an asymmetric coupler design~\cite{Sete2021}, shown in Fig.~\ref{fig:fridge_setup} of Appendix~\ref{app:exp_setup}. The qubits have strong couplings $g_{ic} \sim \SI{60}{\mega\hertz}$ to the coupler, 
 while the direct coupling between the qubits $g_{12}=-\SI{5.5}{\mega\hertz}$ was designed to be small and of opposite sign to reduce the ZZ-interaction in the idle state~\cite{Sete2021}.

\begin{table}[b]
    \centering
    \begin{ruledtabular}
    \begin{tabular}{cccc}

\rule{1cm}{0pt}& $Q_1$ & $Q_2$ & C \\
 $\omega_i/2\pi$ & \SI{4.115}{\giga\hertz} & \SI{3.651}{\giga\hertz} & $\SI{3.7}{} - \SI{6.3}{\giga\hertz}$ \\
 $\alpha_i/2\pi$ & \SI{-261}{\mega\hertz} & \SI{-275}{\mega\hertz} & \SI{-124}{\mega\hertz} \\
$g_{i{\text{c}}}/2\pi$ & \SI{67}{\mega\hertz} & \SI{61}{\mega\hertz} & \\
$g_{12}/2\pi$ & \SI{-5.5}{\mega\hertz} & & \\
$T_1$ & \SI{51\pm 6}{\micro\second} & \SI{125\pm 25}{\micro\second} & \SI{16\pm 7}{\micro\second}\\
$T_2^\text{E}$ & \SI{90\pm 30}{\micro\second} & \SI{30\pm4}{\micro\second} & \\
$T_2^\text{R}$ & \SI{23\pm3}{\micro\second} & \SI{6.6\pm1.4}{\micro\second} & \\
$\mathcal{F}^\text{SQ}$ & \SI{99.93\pm0.01}{\percent}& \SI{99.7\pm0.1}{\percent}& \\

\end{tabular}

    \end{ruledtabular}
    \caption{
        Parameters of the device with two qubits $Q_1, Q_2$ and a tunable coupler $C$.\label{tab:qubit_paramters}
    }
\end{table}

The conditional frequency shift $\xi$ is tunable from $\xi / 2\pi = \SI{20}{\kilo\hertz}$ to   $\xi/ 2\pi> \SI{100}{\mega\hertz}$ via selective energy shifts caused by avoided crossings with tunable coupler states, shown in Fig.~\ref{fig:system_chip}(b).
At the idle position ($\Phi_\text{i}= \Phi_\text{ext} =  0.285\,\Phi_0$), far detuned from any avoided crossing, the eigenstates $\ket{n_1 n_2}$ of the Hamiltonian in Eq.~\eqref{eq:system_ham} are approximately equal to the corresponding bare states labeled $\ket{n_c, n_1n_2}^0$, where $n_1, n_2$ are the excitation numbers for the qubits and $n_c$ the excitation number of the tunable coupler. 
At this point coherence-limited single-qubit gate fidelities $\mathcal{F}^\text{SQ}$ above \SI{99.7}{\percent} are measured in simultaneous randomized benchmarking (RB) experiments (Table~\ref{tab:qubit_paramters}).

Adiabatically lowering the coupler frequency $\omega_c$ by a change in $\Phi_\text{ext}$ leads first to an adiabatic excitation transfer between qubit $Q_1$ and the coupler $C$ via the interaction of the $\ket{0,1\, n_2}^0$ and $\ket{1,0\, n_2}^0$ states at $\Phi_\star = \Phi_\text{ext} = 0.408\,\Phi_0 $ [indicated by $\star$ in Fig.~\ref{fig:system_chip}(b)].
At this point, the energy shift is largely independent of the state of qubit $Q_2$. Only when tuning the frequency $\omega_c$  lower such that the second-excited coupler state $\ket{2,00}^0$ hybridizes with the $\ket{0,11}^0$ and the $\ket{1,01}^0$ state, the conditional frequency shift becomes large. The interaction between these states, indicated by $\hexagon$ in Fig.~\ref{fig:system_chip}(b), is present only in the double-excitation manifold, i.e., it is conditional on $Q_2$ being in the excited state.
At the point $\Phi_\text{f} = \Phi_\text{ext} =  0.456\,\Phi_0$, we observe a conditional frequency shift $\xi/ 2\pi> \SI{75}{\mega\hertz}$, which enables the implementation of conditional phase rotations within nanosecond timescales.

\begin{figure}[t]
    \centering
    \includegraphics{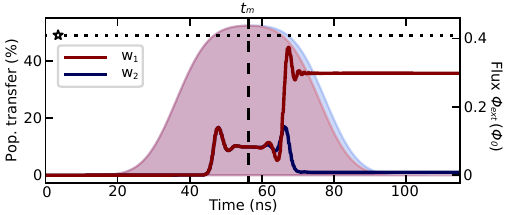}
    \caption{
        Population leakage and retrieval. 
        Simulated population transfer to other states for two Gaussian-square pulses with pulse width $w_1$ and $w_2$ (envelope in the background). The solid lines show the leakage from the $\ket{11}$ state to the $\ket{1,01}$ and $\ket{2,00}$ states.  The dotted line shows the flux amplitude at the avoided crossing $\Phi_\star$.
    \label{fig:amp_width_sweep}
    }
\end{figure}

To measure $\xi$ as a function of external flux, we perform Ramsey measurements interleaved by adiabatic flux pulses, see black circles in Fig.~\ref{fig:system_chip}(c).
The resulting values agree well with a numerical diagonalization of the Hamiltonian in Eq.~\eqref{eq:system_ham}, using the parameters in Tab.~\ref{tab:qubit_paramters}.
In the numerical simulations, the Hamiltonian is truncated to three qubit levels and four coupler levels. 
To implement a controlled-Z operation, the flux pulse must be chosen such that the accumulated conditional phase is $\phi_\text{ZZ} = \int{\xi(t)\text{d}t}=\pi$.
Note that the single-qubit phases $\varphi_1, \varphi_2$ accumulated during the gate operation are corrected using virtual-Z rotations~\cite{McKay2017, Ganzhorn2019}. 

To realize short gate times, the coupler needs to be tuned past the avoided crossings $\Phi_\text{ext}>\Phi_\star$ to an interaction region with a large conditional frequency shift $\xi$.
\label{sec:calibration_errors}
However, to remain in the adiabatic regime the pulse cannot be arbitrarily short~\cite{Collodo2020}, as the adiabatic condition limits the slope of the pulse to  $\text{d}\omega/\text{d} t \ll g^2$ with coupling $g$ between two interacting states. 
Relaxing the adiabaticity condition results in a population transfer between the interacting eigenstates~\cite{Collodo2020,Sung2021, ding2024pulse}.
However, population loss can be minimized by choosing a suitable pulse width, ensuring that any population lost while passing the avoided crossing is recovered when returning to the idle position $\Phi_\text{i}$. This coherent effect is known as Landau-Zener-Stückelberg (LZS) interference~\cite{Shevchenko2010, zener1932non, gefen1987zener} and depends in particular on the relative accumulated phase differences $\Lambda = \int{  \left(\omega_i \left(t \right) - \omega_j\left((t \right) \right) \text{d}t}$ of the populated eigenstates $\ket{i},\ket{j}$.
Using the Landau-Zener model, the remaining population $P(t_e)$ after the return passage is periodic in $\Lambda$, with $P_e = 1- L_\text{m}\cdot \sin^2(\Lambda)$.
The maximal population loss $1 - P_\text{m} = 4 P_m \left(1 - P_m\right)$ is given by the population transferred following a single passage of the avoided crossings ($1 - P_m$).
In Figure~\ref{fig:amp_width_sweep}, two pulses with pulse width $w_1=\SI{39.67}{\nano\second}$ and $w_2=\SI{41.39}{\nano\second}$ are shown, along with a simulation of the population loss from the $\ket{11}$ state to the $\ket{1,01}$ and $\ket{2,00}$ states. This simulation highlights how the evolved phases, influenced by the pulse width, determine whether population loss occurs or is mitigated.

To realize a high-fidelity gate, the errors due to leakage must be minimized while at the same time collecting correct phases. 
For a calibration using a Gaussian-square pulse shape with three pulse shape parameters ($\Phi_f$, $w$, $\tau_r$), we achieve a gate fidelity of \SI{95}{\percent}, as discussed in Appendix~\ref{App:ErrorSources}.
 To further increase the fidelity, more control parameters are needed to simultaneously reduce both leakage errors and phase accumulation errors.

\section{Pulse shape parametrizations}
To realize a gate with minimal error rate, we explore the following pulse parametrizations, which we calibrate using a closed-loop optimization algorithm~\cite{werninghaus2021leakage, kelly2014optimal}: a five-parameter Gaussian-square pulse for a basic implementation of the CZ gate, a piecewise-constant-slope (PiCoS) parametrization with 21 parameters, and a Fourier-component decomposition with eight parameters.

The Gaussian-square pulse serves as a benchmarking baseline, with a minimal parameter set $\param_\text{GS} = \{A, w, \tau_R, \varphi_1, \varphi_2\}$, where $A$ denotes the amplitude scaling, $w$ its pulse width, and $\tau_R$ the rise-time. $\varphi_{1,2}$ are single-qubit Z-rotations for the qubits $Q1$ and $Q_2$. 

\begin{figure}[t]
    \centering
    \includegraphics{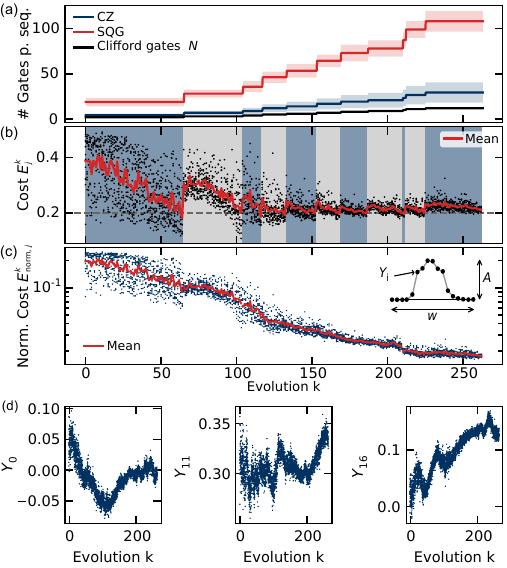}
    \caption{
        Exemplary optimization curves for a PiCoS pulse-shape parametrization. (a) Number of CZ gates and single-qubit gates (SQG) per sequence averaged over the 80 measured randomized sequences. (b) Measured sequence infidelity $E_j^k$ during the optimization. Blue points show the sequence infidelities of each candidate. The red line shows the mean value over the candidate solutions of each evolution. The change in shading indicates increases in cost function sensitivity $S$. (c) Convergence of normalized sequence infidelity $E_{\text{norm}, j}^k = E_j^k / N$ on a logarithmic axis. 
        The inset illustrates a schematic of the PiCoS pulse shape with the Nodes $Y_i$, the amplitude $A$, and the width $w$.
        (d) Evolution of selected nodes $Y_i$ during the optimization of the PiCoS parametrization.
    }
    \label{fig:fig3_3}
\end{figure}

The Piecewise-Constant-Slope (PiCoS) parametrization is a simplification of the widely used piecewise-constant (PWC) parameterization for pulses digitized at the device sampling rate~\cite{khaneja2005, werninghaus2021leakage}.
It is defined by $N_\text{Y}$ amplitudes $Y_i$ with equal spacing $w/N_\text{Y}$ across the pulse width $w$. The $N_\text{Y}$ nodes are interpolated via a constant slope, as shown in the inset of Fig.~\ref{fig:fig3_3}(c).
This approach significantly reduces the $\mathcal{O}(100)$ parameters of the PWC parameterization, which are determined by the sampling rate of the arbitrary waveform generator (in our experiment  $2\,\rm{GS/s}$\footnote{We employ the Zurich Instruments QCCS for the generation of control signals.}) and the pulse duration (typically around \SI{50}{\nano\second}). 
 Additional parameters included in the optimization are a global amplitude scaling $A$, the pulse width $w$, and Z-rotations $\varphi_1, \varphi_2$ applied on the respective qubits. This results in a parameter set $\param_\text{PiCoS} = \{Y_i, A, w, \varphi_1, \varphi_2\}$. While parameters like $A$ and $w$ are redundant with the PiCoS approach, they allow for global adaptations to rescale the pulse, adding significant robustness to the calibration.

A less complex yet similarly versatile pulse shape parametrization is provided by a decomposition of the envelope into a sum of Fourier components
\begin{align}
    \Phi(t)=A  \sum_{n=1}^N \lambda_n\left(1-\cos \left(2 \pi n \frac{t-t_p/2}{\tau_w}\right)\right),
\end{align}
with the maximum amplitude $A$, the width of the pulse shape $w$ and the normalized Fourier-coefficients $\lambda_n$.
This parametrization results in a parameter set $\param_\text{Fourier} = \{A, w, \lambda_n, \varphi_1, \varphi_2\}$.
Using only low-frequency Fourier coefficients smooth pulse shapes can be ensured. Numerical open-loop optimizations indicate that a good balance between the number of control parameters and sufficient controllability can be achieved with five Fourier-coefficients~\cite{Chu2021,glaseroptimal2021}.

 For all three pulse parametrizations, we use the CMA-ES algorithm~\cite{Hansen2016} to simultaneously optimize all parameters, with additional details provided in Appendix~\ref{app:optimizaion}. This algorithm can intrinsically handle the parameter correlations discussed in Appendix~\ref{App:ErrorSources}.
In each optimization iteration (evolution) $k$ the optimizer chooses $P$ parameter-combinations $\{\param_j^{k}, j \in [1, P]\}$ to be probed.
The parameters are sampled from Gaussian distributions, defined via a mean $m^{(k)}$ for each parameter and a covariance matrix.
We utilize the ORBIT protocol to determine a cost $E_j^k$ for each candidate solution $\param_j^{k}$, which reflects the performance of the resulting gate~\cite{kelly2014optimal,werninghaus2021leakage}. The cost measures the qubit population after randomized benchmarking sequences comprising $N$ Clifford gates.
We measure $M=80$ different randomizations that are each averaged over 128 shots.
For every parameter combination $\param_j^{k}$ of the CZ gate, the cost $E_j^k$ is measured.
The weighted average and the covariance of $\param_j^k$ with respect to $E_j^k$ are then used to update the sampling distribution's mean and covariance matrix.
This process is iterated until the distribution has converged to a minimum. 
This makes the optimizer generally well suited for experimental optimization, as it provides robustness to slow drifts in the cost function~\cite{Hansen2016}. 
If possible, all $j$ evaluations should be performed within a single experiment to distribute system fluctuations evenly across all cost function values $E_j^k$.
Furthermore, the classical processing overhead is reduced by using a single experiment per evolution~\cite{werninghaus2021leakage}. 

The measured cost functions $E_j^k$ are influenced by experimental noise, making it essential to maximize the sensitivity $S_k$ during each evolution. The sensitivity of the cost function with respect to changes in the CZ gate error $\epsilon$ is defined as $S_k(N) =  \left(\text{d} E_j^k /\text{d}\epsilon\right)\mathcal{N}^{-1} $, where $\mathcal{N}$ is a normalization constant. The maximum sensitivity depends on the number of Clifford gates $N$ and the average gate errors $\epsilon$ for a given parameter combination $\{\param_j^{k}\}$. 
Although the exact error of the gate is not known during the optimization, the ideal sensitivity $S_k$ can be approximated based on the value of the cost function $E_j^k$, as discussed in more detail in Appendix~\ref{app:cost}.
Generally, the sensitivity $S_k(N)$ remains above $\SI{90}{\percent}$ of its maximum value for cost function values $E_j^k$ ranging between 0.2 and 0.4. Therefore, we adapt the sensitivity by increasing the number of Clifford gates $N$ during the optimization process whenever the mean cost function value across the evolution $\bar{E}^k$ falls below 0.2.

In the CMA-ES algorithm, the cost function can be rescaled during the optimization process as it compares values within each set of parameter combinations $\param_j^{k}$ of a single evolution.  
This contrasts with commonly used optimizers, such as Nelder-Mead~\cite{nelder1965simplex} or L-BFGS~\cite{liu1989limited}, where the cost $E_j^k$ of parameter combinations is added individually and compared with previous evaluations.

An exemplary optimization run for the PiCoS parametrization is shown in Fig.~\ref{fig:fig3_3}. The optimization begins with low sensitivity $S$ and $N=2$ Clifford gates for an initial parameter combination $\{\param_j^{0}\}$, where individual candidate solutions with two-qubit gate errors up to $\epsilon \sim\SI{50}{\percent}$ are measured.
As the optimizer identifies combinations with higher fidelities, the sensitivity $S$ and, therefore, the number of gates is incremented in steps of $N \rightarrow N+1$ to $N=12$ Clifford gates [Fig.~\ref{fig:fig3_3}(a)]. 
This stepwise increase allows for the separation of small variations between the high-fidelity candidate parameters, as the mean cost per evolution $\bar{E}^k$ remains between 0.2 and 0.4 [Fig.~\ref{fig:fig3_3}(b)].  
    
The optimization of the PiCoS pulse with 21 parameters converges after 265 evolutions and a total time of \SI{12}{\hour} to a gate error of $\epsilon^\text{PiCoS} = \SI{0.25\pm 0.09}{\percent}$.
The optimization is terminated if $\bar{E}^k$ remains steady over multiple evolutions $k$.
The final result of the optimization is chosen as the parametrization $\param_j^{k}$ with the lowest $E_j^k$ within all evolutions since the last sensitivity update.
  To illustrate the convergence of the protocol, we compute the normalized Clifford infidelity $E_{\text{norm}, j}^k = E_j^k / N$  by dividing the measured sequence infidelity $E_j^k$ by the number of Cliffords per sequence. After 200 evolutions, $E_{\text{norm}, j}^k$ is reduced by an order of magnitude, as shown is shown in  Fig.~\ref{fig:fig3_3}(c). A select set of amplitude values $Y_i$ is shown in Fig.~\ref{fig:fig3_3}(d).

The required number of evolutions $k_\text{max}$ and parameter combinations per evolution $P$ depend on the number of pulse-shape parameters, as further discussed in  Appendix~\ref{app:CMA-dimensionality}.

\begin{figure*}[tb]
    \centering
    \includegraphics[]{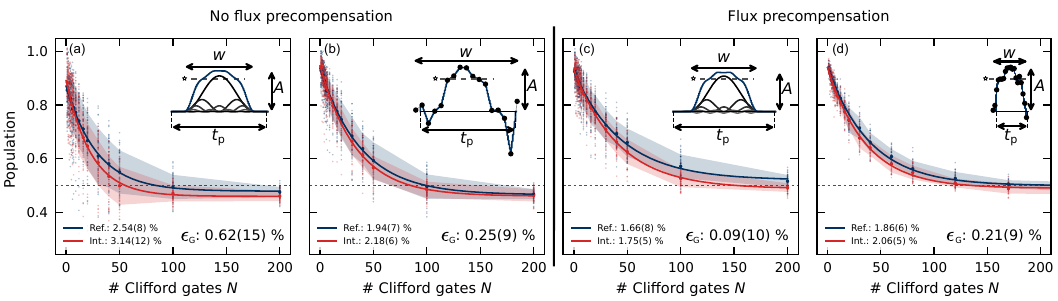}
    \caption{
        Optimized pulse-shape parametrizations and interleaved randomized benchmarking results comparing (a, c) the Fourier-series parametrization and (b, d) the PiCoS parametrization, with gate errors denoted as $\epsilon$. Panels (a) and (b) display the results before applying pulse pre-compensation, while panels (c) and (d) show the results after the implementation of pre-compensation.
    }
    \label{fig:comparison}
\end{figure*}

\section{High-fidelity CZ gate optimization}
We apply the optimization algorithm described above to different pulse parametrizations.
Initially, the pulse shapes are coarsely calibrated in amplitude,  pulse width, and single-qubit phases  $\varphi_{1}$, $\varphi_{2}$ ensuring an error of $\lesssim \SI{10}{\percent}$.
The gates are then optimized using the optimization procedure described above.
Following the optimization, we evaluate the gate error $\epsilon$ using interleaved randomized benchmarking~\cite{magesan2012efficient}.
The performance of the different pulse implementations, along with their respective pulse durations, is summarized in Tab.~\ref{tab:performance}.

The calibrated Gaussian-square pulse parametrization with a pulse width $w=60~\rm{ns}$ results in a gate error of $\epsilon^\text{G-Sq}=\SI{4.79(19)}{\percent}$. 
This high error indicates that the gate is strongly limited by coherent processes, suggesting that this parametrization is not well-suited for achieving high-fidelity coherence-limited gates.
In contrast, gates with the Fourier-series parametrization with five Fourier-components and a pulse width $w=\SI{60}{\nano\second}$ achieve an optimized gate infidelity of $\epsilon^\text{Fourier}= \SI{0.62\pm0.15}{\percent}$, see Fig.~\ref{fig:comparison}(a).
This is a significant improvement in comparison to the Gaussian-square pulse and the coarse calibration.
 This optimization converges after 80 evolutions, requiring a calibration time of \SI{142}{\minute}. 
With the PiCoS pulse parametrization, the gate error is further reduced by a factor of three to $\epsilon^\text{PiCoS}= \SI{0.25\pm0.09}{\percent}$, see Fig.~\ref{fig:comparison}(b).
Moreover, interleaved randomized benchmarking measurements for the PiCoS pulses show a lower spread  between randomizations compared to the Fourier-series pulse, indicating a reduction of unitary errors~\cite{ball2016effect}.

Comparing the Fourier-series and PiCoS pulses the pulse shape is similar, with a distinct slowdown of the trajectories upon entering the interaction region near~$\star$. 
This aligns with the expectation that reducing the  slope at an avoided crossing helps to minimize leakage.
The Fourier-series pulse exhibits signs of leakage out of the qubit subspace, indicated by the population deviating from the fully depolarized value in the limit of large $N_C$~\cite{wood2018quantification,wallman2016robust,werninghaus2021leakage}.
 In contrast, the PiCoS pulse mitigates leakage, due to its additional degrees of freedom. Most leakage would occur into the coupler states as confirmed by numerical simulations. 
 However, since the coupler states are not directly measurable on the current device, an experimental verification of the leakage channels remains outside of the scope of this work.

The optimization of the PiCoS pulse also results in asymmetric shoulders and introduces negative flux values in place of the zero-amplitude regions, see Fig.~\ref{fig:comparison}(b).
This suggests that the optimization routine seeks to compensate for distortion effects in the control line. 
However, due to the slow response time on the order of microseconds, the routine cannot fully eliminate all distortion effects within the pulse duration.
Therefore, memory effects influencing subsequent pulses in the experimental sequence remain uncorrected~\cite{Rol2019}.
To address these limitations and further enhance gate quality, we characterize the line transfer function using an extension of the cryoscope protocol described in \cite{Rol2020}. 
 The extension is adapted to our architecture, which does not allow direct readout of the coupling element.
 Instead, we utilize the interaction between the qubit and the tunable coupling element, as described in more detail in Appendix~\ref{app:flux_distortion}. The characterization reveals significant low-pass filtering effects that limit the fidelity of consecutive CZ gates. To counteract the signal distortion, we systematically correct the transfer function by employing on-device signal pre-distortion using the HDAWG from Zurich Instruments~\cite{Zurichinstrumentsag2022}. With this, we suppress  low-pass filtering, reducing the characteristic time scales from \SI{2.5}{\micro\second} to \SI{10}{\nano\second}. 
 The optimizer can then adapt the pulse shapes more effectively, fully compensating for remaining distortions within the pulse duration.

Reoptimization of Fourier-series pulses with  \SI{64}{\nano\second}, now incorporating pre-distortions, achieves significantly reduced gate errors of $\epsilon^\text{Fourier, C} = \SI{0.09 \pm 0.1}{\percent}$, see Fig.~\ref{fig:comparison}(c). 
Repeating the procedure for the PiCoS pulse reveals that, while the gate fidelity has reached its optimal level, there remains potential to reduce the gate duration.
 In fact, for pulses with a length of \SI{20}{\nano\second} followed by a \SI{12}{\nano\second} buffer period of zero signal amplitude, we achieve a gate error as low as $\epsilon^\text{PiCoS, C} = \SI{0.21\pm 0.09}{\percent}$, as shown in Fig~\ref{fig:comparison}(d). 
 Further reducing the pulse length, however, leads to increased gate errors, which we attribute primarily to  line distortions occurring on similar timescales.
 For the presented pulse, we observe minimal indications of leakage from the interleaved randomized benchmarking experiment, particularly when compared to the Fourier-series pulse.

\begin{table}[b]
\begin{ruledtabular}
    \begin{tabular}{lcccll}
   &$n_\text{p}$  & $t_p$ & $t_b$ & $\epsilon$ \\
\hline
GS & 5& $50 \mathrm{~ns}$ & \SI{50}{\nano\second} & $4.79(19) \%$ \\
\hline
PiCoS  & 8& $60 \mathrm{~ns}$ & \SI{0}{\nano\second}  & $0.52(15) \%$ \\
\hline
Fourier & 21   & $60 \mathrm{~ns}$ & \SI{0}{\nano\second}  & $0.25(9) \%$  \\
\hline
PiCoS, C & 8& $64 \mathrm{~ns}$ & \SI{0}{\nano\second}  & $0.09(10) \%$    \\
\hline
Fourier, C & 21   & $20 \mathrm{~ns}$ & \SI{12}{\nano\second} & $0.21(9) \%$\\
\end{tabular}
\end{ruledtabular}
    \caption{Characteristics of optimized gates. Performance of discussed gates Gaussian-square (GS), Fourier-series, and PiCoS-pulse, with number of optimized parameters $n_p$, fixed pulse length $t_p$, CZ gate error $\mathcal{\epsilon}$ and applied precompensation (Fourier, C) and (PiCoS, C)}.\label{tab:performance}
\end{table}
\section{Discussion and Conclusion}

We have demonstrated a closed-loop optimization of parametrized pulses in combination with a signal pre-distortion to correct on time scales exceeding the gate duration. This approach achieves two-qubit gate fidelities around \SI{99.9}{\percent} in less than \SI{64}{\nano\second}, a significant improvement compared to Gaussian-square pulses, which yield error rates of around~\SI{5}{\percent} but can be optimized using conventional tune-up methods based-on individual parameter calibrations.
Our approach of a closed-loop CMA-ES-based simultaneous parameter optimization in conjunction with increased complexity pulse shapes, i.e., Fourier-Series and PiCoS pulses, enables gate errors to be reduced below $\SI{0.3}{\percent}$.
Notably, we have demonstrated that the optimized PiCoS parameterization allows us to reduce leakage and intrinsically correct flux-line distortions.

In addition, optimization results have been utilized to identify systematic errors. In particular, flux-line distortions are reflected in the pulse shapes after the optimization routine has converged to a high-fidelity gate. 
This inference of system properties through black-box optimal control, followed by the subsequent compensation of systematic deviations enables the realization of control pulses with reduced complexity, a cornerstone of optimal control theory~\cite{Machnes2018,werninghaus2021leakage,Wittler2021}. System identification, as demonstrated in this work, will facilitate more accurate open-loop optimal control and provides an improved understanding of the system parameters.

As part of the systematic distortion mitigation, we extended the cryoscope protocol~\cite{Rol2020} to indirectly measure flux-line distortions via non-tunable elements. By inverting the system transfer functions directly on the classical control hardware~\cite{Zurichinstrumentsag2022}, we compensate for these distortions. This approach suppresses long-term memory effects on the scale of the experimental sequences, which allows us to reach average gate fidelities above \SI{99.9}{\percent}.

Furthermore, the closed-loop optimal control scheme shows promise for quantum processor tune-up application. For this the required time per optimization has to be further reduced. We have shown that this can be achieved through system identification, reducing the optimization time from \SI{12}{\hour} to \SI{4.5}{\hour} by using a pulse shape with less complexity. 
We have analyzed the primary factors contributing to optimization runtime,  including classical processing time and passive reset, as detailed in Appendix~\ref{app:walltime}.
There, we outline strategies to decrease the runtime by more than a factor of 100. 

This optimization method is also well-suited for recalibration tasks, where meta-parameters such as previous parameter spreads and covariance matrices can be retained. This allows the recalibration process to  convergence rapidly, especially when only minor adjustments to the optimal parameters are required.
This approach could ensure high system uptime by interleaving optimization evolutions with algorithmic tasks, thereby tracking and compensating for small parameter drifts in real time. Such a strategy would preserve high-fidelity operations throughout the algorithm runtime. 

\section{Acknowledgements}
We acknowledge financial support from the German Federal Ministry of Education and Research via the funding program \textit{Quantum Technologies – From Basic Research to the Market} under contract number 13N15680 \textit{\mbox{GeQCoS}} and 13N16188 \textit{MUNIQC-SC}; the Deutsche Forschungsgemeinschaft (DFG, German Research Foundation) via project number FI2549/1-1 and Germany's Excellence Strategy EXC-2111-390814868 \textit{MCQST}; and the European Union's Horizon research and innovation program through the projects \textit{OpenSuperQPlus100} (Grant-Nr. 955479) and \textit{MOlecular Quantum Simulations (MOQS)} (Grant-Nr. 955479). We also acknowledge support from the the EU MSCA Cofund \textit{International, Interdisciplinary, and Intersectoral Doctoral Program in Quantum Science and Technologies (QUSTEC)} (Grant-Nr. 847471). This research is part of the \textit{Munich Quantum Valley}, supported by the Bavarian state government  with funds from the Hightech Agenda Bayern Plus.

\bibliography{bibliography}
\clearpage
\appendix

\section{Experimental setup}
\label{app:exp_setup}
\begin{figure}[b]
    \includegraphics{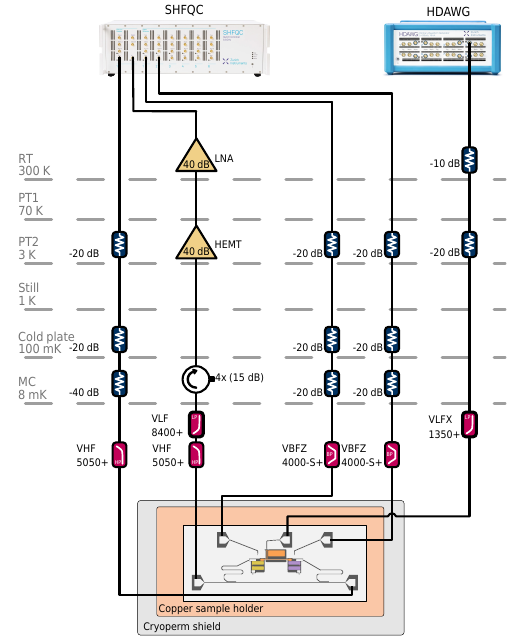}
    \caption{Experimental setup and schematic of the wiring for the superconducting qubit experiment.\label{fig:fridge_setup}}
\end{figure}

The qubits are controlled and measured using a \textit{Super High Frequency Qubit Controller} (SHFQC) from Zurich Instruments. A schematic of the experimental setup is shown in Fig~\ref{fig:fridge_setup}.
We use charge-coupled direct drive lines on the chip for qubit control. The readout resonators are coupled to a common feedline.
 The feedline is attenuated by \SI{-80}{\dB} distributed over the temperature stages and filtered with a \SI{5.5}{\GHz} high pass filter (VHF-5050+).
The drive lines are attenuated by \SI{-60}{\dB} distributed over the different temperature stages and filtered with a \SI{3.5}{\GHz} to \SI{4.5}{\giga\hertz} band pass filter (VBFZ-4000-S+).

We use two $4-\SI{12}{\GHz}$ isolators on the output line.
The output signal is amplified by a $4-\SI{8}{\GHz}$ \SI{40}{\dB} HEMT cryogenic low-noise amplifier (LNF-LNC4\_8G) thermalized at the \SI{3}{\kelvin} stage.
The signal is amplified by a \SI{40}{\dB} low-noise room temperature amplifier (DBLNA104000800A), before being digitized at the SHFQC input.
The flux offset and drive to the coupler is generated using one channel of a Zurich Instruments HDAWG. The flux line is attenuated at room temperature by \SI{-10}{\dB} and by \SI{-20}{\dB} at the \SI{3}{\kelvin} stage and filtered using a \SI{1.35}{\GHz} low-pass filter (VLFX-1350+).

\section{ZZ-Phase and leakage}\label{App:ErrorSources}
\begin{figure*}[tb]
    \centering
    \includegraphics[]{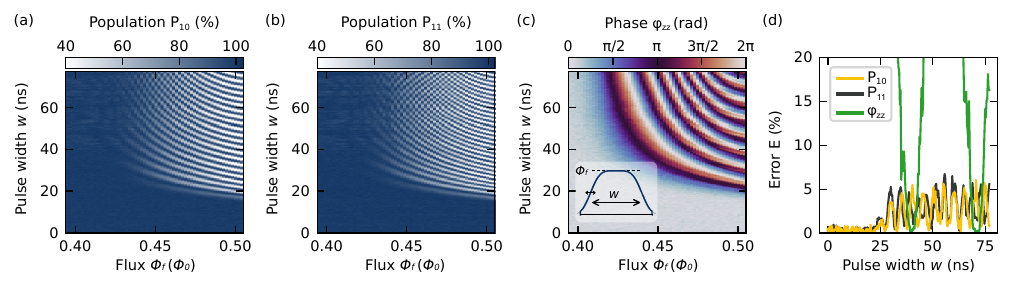}
    \caption{
    Leakage induced by the CPHASE gate pulse measured from the (a) $\ket{10}$ and (b) $\ket{11}$ states, alongside the (c) accumulated ZZ-phase $\varphi_\text{ZZ}$. Measurements are shown as functions of pulse width $w$ and amplitude $\Phi_\text{f}$. (d) Error rates $E$ for population loss from the $\ket{11}$ ($1- P_{11}$) and $\ket{10}$ ($1 - P_{10}$) states, along with the accumulated ZZ-phase $\phi_\text{ZZ}$, at a fixed amplitude of $\Phi_\text{f} = 0.44 \, \Phi_0$. Note that the error is truncated at \SI{20}{\percent}.
    }
    \label{fig:leakage_error_zoom}
\end{figure*}

We illustrate the effect of coherent errors emerging from non-ideal pulse shapes and their correlated dependence on pulse-shape parameters using simple pulse parametrizations. 
As discussed in the main text the hybridization with the avoided crossings may lead to population losses.
We tune the flux of the coupler, and thereby the coupler frequency, using a Gaussian-square pulse defined as
\begin{align}
      \Phi(t) = \Phi_\text{f} \cdot 
       \prod_{\pm} \frac{1}{2}\left(1 + \erf\left(\tfrac{\pm t+  w/2}{\tau_R}\right)\right),
\end{align}
where $w$ is the pulse width, $\Phi_\text{f}$ is the maximal flux value, and $\tau_R$ is the rise time and $\erf$ is the Gauss error function. An illustration of the parametrization is shown in the inset of Fig.~\ref{fig:leakage_error_zoom}(c).
We measure the population $P_{n_1n_2}$  after applying a flux pulse to the coupler, with the system initialized in one of the computational basis states $\ket{\psi}_i = \ket{P_{n_1n_2}}$.
Figs.~\ref{fig:leakage_error_zoom}(a,b) depict the final population of $\ket{\psi}_2=\ket{10}$ and $\ket{\psi}_3=\ket{11}$ as functions of the pulse width $w$ and the flux amplitude $\Phi_\text{f}$.
For these states, we observe a population transfer indicating stronger interactions between the coupler and the qubit states for  increasing flux pulse amplitude $\Phi_\text{f}$.
A significant increase in population transfer occurs when the system enters the regime beyond the avoided crossing at $\Phi_\text{f}> 0.43\Phi_0$ (see Fig.\ref{fig:system_chip}(b)).
The population transfer exhibits coherent oscillations with respect to the pulse width $w$, consistent with Landau-Zener-Stückelberg (LZS) oscillations.
For short gate times $w<\SI{20}{\nano\second}$, no leakage is observed, as the maximum pulse amplitude is constrained by the pulse rise time.
For the states $\ket{\psi}_0=\ket{00}$ and $\ket{\psi}_1=\ket{01}$, the population transfer is within the measurement errors and is therefore not shown.

Achieving a high-fidelity gate requires carefully selecting the amplitude $\Phi_f$ and pulse width $w$ to minimize leakage while ensuring the correct acquired ZZ-phase $\phi_\text{ZZ}$.
The ZZ-phase $\phi_\text{ZZ}$ is measured via Ramsey-type experiments, following a protocol similar to that in~\cite{Ganzhorn2019}.
The control qubit is initialized in either $\ket{0}$ or $\ket{1}$, while the target qubit is set to the $\ket{-Y}$ state using an $R_{\pi/2}^{X}$ rotation. 
The CZ gate, consisting of the flux pulse on the coupler and Z-rotations on the qubits, is then applied.
A final $R_{\pi /2}^{\theta}$ pulse maps the evolved state back to the z-axis for measurement.
The rotation axis angle $\theta$ is swept over ten points  from $-1.1\pi$ to $1.1\pi$, and the resulting population is fitted to a cosine function to extract the acquired phase.
The ZZ-phase $\phi_\text{ZZ}$ is determined as the difference between the acquired phases with the control qubit in $\ket{1}$ and $\ket{0}$.

The errors in leakage and ZZ-phase $\phi_\text{ZZ}$ are estimated using the average gate fidelity~\cite{nielsen2002simple}.
The phase mismatch error was derived to be 
\begin{align}
E_\varphi = 1 - \left( \frac{2(d-1)}{d^2 + d}\cos{\delta\varphi}  + \frac{d(d-1)+2}{d^2+d}\right),
\end{align}
and population loss error is 
\begin{align}
    E_P = 1 -\frac{1}{(d^2 + d)}\left(d + \left|d - \sum{1 - P_{n_1n_2}}\right|^2\right),
\end{align}
 where $d$  is the Hilbert space dimension.
Simultaneously minimizing both errors is essential for a high-fidelity pulse. As shown in Fig.~\ref{fig:leakage_error_zoom}(d), for a slice at $\Phi_f = 0.44\,\Phi_0$, the estimated errors exhibit unsynchronized oscillations making it difficult to identify parameters for a common minimum error. 
As discussed in the main text, employing more complex pulse shapes with additional parameters help to mitigate both errors simultaneously.

\section{Optimization} \label{app:optimizaion}
This appendix outlines the optimization process, including the ORBIT cost function and its sensitivity, the algorithm with its parameters, runtime analysis, and convergence time as a function of the parameter count.
\subsection{ORBIT cost function}\label{app:cost}

\begin{figure}[b]
    \centering
    \includegraphics[]{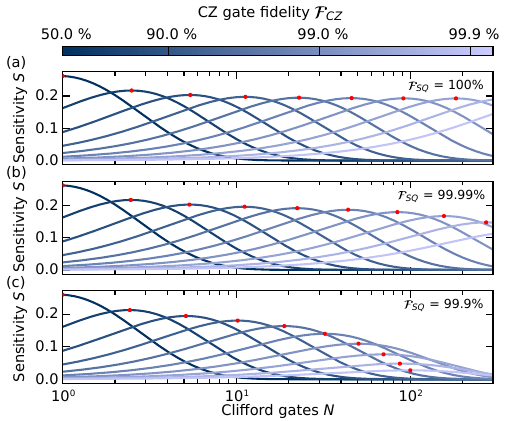}
    \caption{
        Sensitivity $S$ of the Clifford sequence to fidelity changes plotted as a function of the Clifford gate count $N$ using Eq.~\eqref{eq:sensitivity} for CZ gate errors $\epsilon_\text{CZ} \in[0.5,0.001]$. Different CZ gate errors are indicated by the color shading. For single-qubit gate fidelities we choose (a) $F_\text{SQ}=100\,\% $ (b) $F_\text{SQ}=99.99\,\% $ and (c) $F_\text{SQ}=99.9\,\% $, respectively. Red dots indicate the points of maximal sensitivity.}
    \label{fig:sensitivity}
\end{figure}
 The cost function for the optimization is derived from randomized benchmarking sequences of fixed length~\cite{kelly2014optimal}.
The sequence fidelity of the Clifford-based cost function depends on the number of Clifford gates $N$ and follows the form \mbox{$\mathcal{F}(N) = A \left(\mathcal{F}_{\mathcal{C}}\right)^N + B$}~\cite{magesan2012efficient}.
The fidelity per  Clifford gate, $\mathcal{F}_{\mathcal{C}}$, is approximated as $\mathcal{F}_{\mathcal{C}} = \left(\mathcal{F}_\text{SQ}\right)^{\bar{n}_\text{SQ}} \cdot \left(\mathcal{F}_\text{CZ}\right)^{\bar{n}_\text{CZ}}$, where $\mathcal{F}_\text{SQ}$ and $\mathcal{F}_\text{CZ}$ are the single and two-qubit gate fidelities, respectively, and $\bar{n}_\text{SQ}=8.5$ and $\bar{n}_\text{CZ}=1.5$ represent the average number of gates per Clifford. This approximation is based on a Clifford gate decomposition that includes $X_{\pi}$, $Y_{\pi}$, $X_{\pi/2}$, $Y_{\pi/2}$, $X_{-\pi}$, $Y_{-\pi}$, $X_{-\pi/2}$, $Y_{-\pi/2}$ and CZ gates, under the assumption of uncorrelated errors.

The sensitivity $S$ of the sequence fidelity $\mathcal{F}(N)$ to parameter changes depends on the number of Clifford gates and the fidelity of the individual gates in the sequence.
The sensitivity $S$ to changes in the CZ gate fidelity, normalized by the CZ gate error, is evaluated with 
\begin{align}
S=\frac{1}{1-\mathcal{F}_\text{CZ}}\frac{\text{d} \mathcal{F}}{\text{d} \mathcal{F}_\text{CZ}} \label{eq:sensitivity}    
\end{align}
and is shown in Fig.~\ref{fig:sensitivity} for various CZ and single-qubit gate fidelities.
For CZ gates with $\mathcal{F}_\text{CZ} \approx \SI{50}{\percent}$, optimal sensitivity $S$ is achieved using short sequences of $N=2$ Clifford gates.
As gate fidelities improve, achieving maximal sensitivity requires longer sequences, with $N$ determined by $N  = -1/\log( \mathcal{F}_\mathcal{C})$. 

Because single-qubit gates are part of the Clifford decomposition, their fidelity $\mathcal{F}_\text{SQ}$ also influences the optimal choice of $N$.
As shown in Fig~\ref{fig:sensitivity}(b,c) errors in the single-qubit gates reduce the optimal $N$ and weaken the sensitivity to changes in $\mathcal{F}_\text{CZ}$.

\subsection{Optimization algorithm}\label{app:alg}

\begin{algorithm}[tb]
    \caption{Optimization}\label{alg:optimization}
    \begin{algorithmic}[1]
        \Function{optimization}{}:
        \State optimizer $\gets$ \Call{CMAES}{}
        \State $N\gets= 1$
        \State seqs $\gets$ \Call{Get\_Sequences}{length=N}
        \State $k\gets= 0$
        \Repeat
        \State $\param_j^k$ $\gets$ optimizer.\Call{Ask}{P}
        \State experiments $\gets$ []
        \For{$s_j^k$ $\in_j$ $\param_j^k$}
        \State experiments $\overset{+}{\leftarrow}$ \Call{compile}{seqs, $s_j^k$}
        \EndFor
        \State results $\gets$ \Call{execute}{experiments}
        \State $E_j^k$ $\gets$ \Call{process}{results}
        \State optimizer.\Call{Tell}{$E_j^k$}
        \If{\Call{mean$_j$}{$E_j^k$} $<$ 0.2}
        \State N $\gets$ N + 1
        \State sequence $\gets$ \Call{Get\_Sequence}{length=N}
        \EndIf
        \State $k$ $\gets$ $k + 1$
        \Until{stop}
        \EndFunction
    \end{algorithmic}
\end{algorithm}

\begin{figure}
    \centering
    \includegraphics{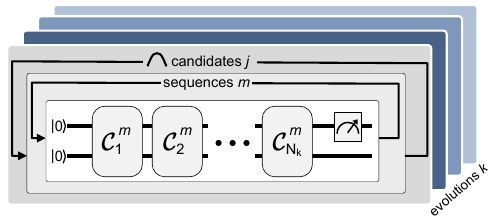}
    \caption{Illustration of the execution flow of the quantum circuits. The process involves $N$ Clifford gates $\mathcal{C}_i^{m}$ followed by a measurement, iterated over multiple shots. This procedure is repeated for $M$ randomized sequences, indexed by $m$. Additionally, the process is repeated for $J$ candidate solutions, indexed by $j$, in order to evaluate the parameter set $\param_j^k$ in each evolution $k$.}
    \label{fig:loops}
\end{figure}

The pulse optimization procedure is carried out using the CMA-ES optimizer~\cite{Hansen2016}, as outlined in Algorithm ~\ref{alg:optimization}.
First, the CMA-ES optimizer is initialized with the parameter set $\param_j^k$ to be optimized, along with the respective hyperparameters, such as their initial values and search range. The parameters are normalized based on the expected optimization range. The choice of the initial parameter spread $\sigma$ strikes a balance between global exploration and local exploitation with its exact value depending on the particular pulse optimization.

A batch of candidate solutions $\param_j^k$ with population size $P$ is sampled by the optimizer based on a Gaussian distribution.
The cost value $E_j^k$ is computed for each candidate solution by applying the gate parametrized according to  $\param_j^k$ in the decomposition of the Clifford sequences (Fig.~\ref{fig:loops}).
We prepare $M=80$ Clifford sequences, each consisting of $N$ Clifford gates. 
The last Clifford gate is chosen to invert the operation of the previous gates, so that the entire sequence effectively implements a pseudo-identity operation. 
For interleaved randomized benchmarking~\cite{magesan2012efficient} half of the $M$ sequences are interleaved with two CZ gates, placed between each decomposed Clifford gate. The set of Clifford sequences remains unchanged for all candidate solutions probed with the same $N$.

The cost value $E_j^k$ is calculated as the average final ground state population of $Q_1$ over $M$ with each 128 repetitions (shots). 
We use LabOneQ~\cite{zhinstlaboneq2024} to compile the instruction set for the control devices based on the candidate solution parameters $\param_j^{k}$, and the Clifford sequences before execution on the control hardware.
If the mean cost $\bar{E}^k = \text{mean}_j(E_j^k)$ of the evolution $k$ falls below 0.2, the number of Clifford gates per sequence is increased by one to enhance the sensitivity $S$ of the cost function. The execution is terminated when the mean cost  $\bar{E}^k$ remains steady over multiple evolutions $k$.
Alternatively, to improve the signal-to-noise ratio (SNR) the number of Clifford sequences $M$ itself or the number of experimental repetitions can be increased. However, this comes at the expense of a longer runtime per evolution due to the increased number of measurements, while maintaining a constant trigger rate.

\subsection{Optimization run time}\label{app:walltime}
\begin{figure}[t]
    \centering
    \includegraphics[]{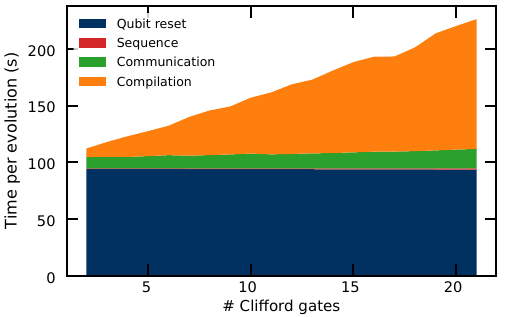}
    \caption{
       Breakdown of the time required per evolution during the optimization process, categorized into qubit reset, sequence execution, device communication, and compilation duration, as a function of the number of Clifford gates in the executed sequences.
    }
    \label{fig:Duration_wall}
\end{figure}

The optimization duration is primarily determined by the number of evolutions $k_\text{max}$ and the time required per optimization step, i.e., per evolution $k$.
The duration is limited from below by the time needed to execute the quantum circuits, including single- and two-qubit gates, measurements and reset operations multiplied by the number of experimental repetitions and sequence randomizations.
To evaluate the overhead caused by device instruction preparation and communication we consider an optimization with a repetition rate of $\SI{1.4}{\kilo\hertz}$, $M=80$ sequence randomizations, $128$ single-shot repetitions, and 13 candidate solutions, 
resulting in a measurement time of 93 seconds for one evolution $k$ (blue + red bar in Fig.~\ref{fig:Duration_wall}).  
In our experiment, the repetition rate is primarily bound by the time required for thermal qubit reset (blue bar in Fig.~\ref{fig:Duration_wall}), while the actual sequence duration, including measurements and pulses, is almost negligible (red bar in Fig.~\ref{fig:Duration_wall}).

The communication time required for control-device programming and data transfers adds a constant offset to the time spent per evolution. This time scales only minimally with sequence complexity and, therefore, the number of Clifford gates per sequence (green area in Fig~\ref{fig:Duration_wall}).
However, the compilation of waveforms and instructions in LabOneQ~\cite{zhinstlaboneq2024} for device control scales linearly with the number of Clifford gates per sequence (orange area in Fig~\ref{fig:Duration_wall}). 

The runtime can in the future be reduced by employing active reset mechanisms~\cite{riste2012initialization, magnard2018fast,  reed2010fast, geerlings2013demonstrating}. 
Assuming measurement-based active reset, the on-device measurement time, including reset pulses, corresponds to $\SI{0.85}{\second}$ per evolution for the given number of repetitions, meaning the total measurement time of the presented PiCoS pulse optimization would be under four minutes.
However, due to the limited readout fidelities, which can be improved by an increased resonator coupling, this was beyond the capabilities of the device under investigation.

Furthermore, a full compilation of instructions and waveforms is used.
If only minimal updates (atomic waveforms + virtual-Z rotation values) were required, the classical processing time could be drastically reduced. 
By using efficient processing methods, the classical processing time could in the future be reduced from  \SI{120}{\second} to \SI{0.72 \pm 0.01}{\second}. 
This has been validated in a proxy setup with reduced experimental complexity and now needs to be implemented at the full algorithmic scale.
\subsection{Convergence speed}\label{app:CMA-dimensionality}
\begin{figure}[t]
    \centering
    \includegraphics[]{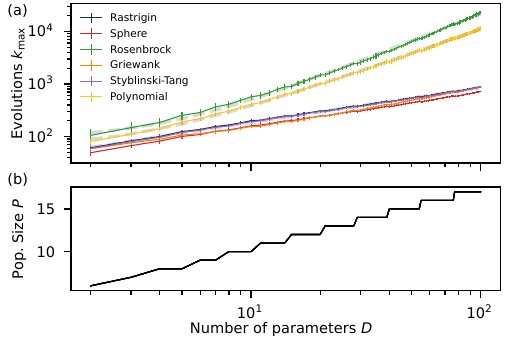}
    \caption{
    (a) Scaling of the required number of optimization evolutions $k_\text{max}$ with the number of parameters $D$ for various test functions, including Rastrigin, Sphere, Rosenbrock, Griewank, Styblinski-Tang, and Polynomial. (b) Population size $P_k$ per evolution as a function of $D$, following the logarithmic scaling $P_k(N_\text{p}) = 4 + 4 \log(N_\text{p})$
    \label{fig:ScalingOfEvolutions}
    }
\end{figure}

Increasing the complexity of the pulse shape can yield higher-fidelity solutions. 
However, the increased number of parameters to be optimized also affects the evolutions $k_\text{max}$ required for convergence, thereby impacting the overall optimization duration.
To guide hyperparameter selection, we examine the scaling behavior of the algorithm's required iterations using several standard optimization problems.
 For the CMA-ES optimizer, we numerically optimize over test functions with added normally distributed noise. 
We evaluate the number of evolutions $k_\text{max}$ required until the variance of the cost function results $E_j^k$ across the candidate solutions $j$ becomes smaller than the variance of the added noise.
For the evaluation of the performance of optimization algorithms, a variety of test functions are commonly used as benchmarks~\cite{PracticalGeneticAlgorithms, back1996evolutionary}. Using such test functions, we observe that the number of evolutions  follows a power law dependence $k_\text{max}\sim{D}^\nu+k_1$, where $D$ is the dimension of the parameter space and $k_1$ is an offset according to the single parameter cases, see Fig.~\ref{fig:ScalingOfEvolutions}.
 For the Spherical, Styblinski-Tang, and Rastrigin test functions, the increase is sub-linear with $\nu\approx0.65$.
 In contrast, for the Rosenbrock and a polynomial function, the scaling factors are $\nu=1.7$ and $\nu=1.5$, respectively.
These scaling factors align with the experimental observations.
We observe an approximately linear relationship between the required evolutions $k_\text{max}$ and the number of parameters in the pulse optimization, using different  parametrizations.

The population size $P$ of the candidate solutions evaluated per evolution $k$ is also influenced by the number of parameters $N_\text{p} = |\param|$ and is determined as $P(N_\text{p}) = 4 + 4 \log(N_\text{p})$~\cite{Hansen2016}, resulting in a logarithmic scaling of the required candidate points, as shown in Fig.~\ref{fig:ScalingOfEvolutions}(b).

\section{Flux distortion}
\label{app:flux_distortion}
\subsection{Signal line characterization \label{app:flux_distortion_correction}}

The transients of the flux pulses are characterized by an extension of the \textit{cryoscope} protocol described in~\cite{Rol2020}.
The original protocol determines the response of the flux line by measuring the change in the frequency of the qubit connected to it.

In this work, however, the response of the coupler instead of the qubit has to be measured. Since the coupler is not directly connected to a readout line, the protocol has been extended, and the response is measured via the coupled fixed-frequency qubit  $Q_1$. 
The eigenstate of the fixed-frequency qubit $Q_1$  experiences a frequency shift due to the hybridization with the coupler state (see Fig.~\ref{fig:system_chip} in the main text).

The phase shift of the qubit state is measured using a Ramsey-type experiment. 
Qubit $Q_1$ is initialized in the ground state and then placed into a superposition state by an $R^X_{\pi/2}$ pulse.
A nominally rectangular flux pulse is applied via the arbitrary waveform generator (Zurich Instruments HDAWG), tuning the coupler's frequency close to the fixed-frequency qubit's frequency. The qubit frequency $\tilde{\omega}_1$ is shifted by the interaction with $\omega_c(\Phi)$.
A final $R^{\theta}_{\pi/2}$ pulse with a variable rotation axis angle $\theta$ is applied after a fixed interval following the initial $R^X_{\pi/2}$ gate.

\begin{figure}[t]
    \centering
    \includegraphics[]{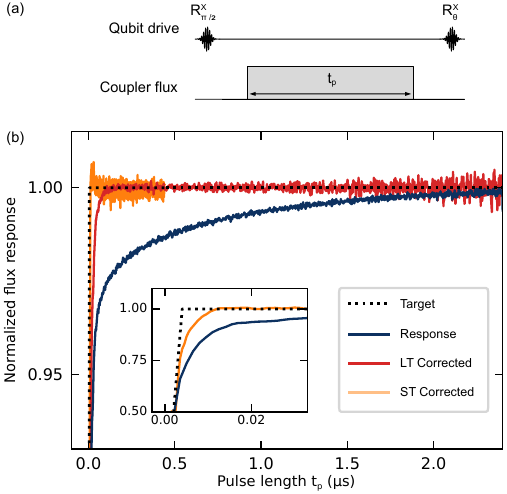}
    \caption{
        Measurement of the coupler flux response. 
        (a) Schematic of the sequence used to measure the response of the flux to the applied pulse. The qubit drive sequence is depicted, including the initial $R_{\pi/2}^X$ rotation, the coupler flux pulse of duration $t_p$, and the final $R_{\pi/2}^\theta$ rotation 
        (b) Measured flux responses. 
        The response is plotted as a function of time, normalized to the calibrated nominal amplitude.
        The uncorrected flux response is shown in blue, showing significant deviation from the target pulse due to distortions over various time scales.
        The flux response after applying long-time scale corrections (red curve) using two exponential IIR filters, improving the alignment with the target pulse.
        The fully corrected flux-response with four exponential IIR and FIR filter correction.
        The black dashed line shows the nominal pulse send to AWG.
        The inset provides a zoomed-in view of the initial \SI{30}{\nano\second} of the response.}
    \label{fig:Distortion}
\end{figure}

In our experiment, we use nine linearly spaced axis angles $\theta$ in the range $[0,2\pi]$, which allows us to extract the phase shift via a cosine function fit. With the individual phase fits per time-step, the data is robust against fluctuations of the coherence time and population loss of the qubit. The flux pulse length $t_p$ is varied from $\SI{0}{\micro\second}$ to $\SI{2.5}{\micro\second}$, as shown in Fig.~\ref{fig:Distortion}(a).
The effective qubit frequency change as a function of the flux pulse length $t_p$ is estimated by the numerical derivative of the phases with respect to the flux pulse length $t_p$.
 Suddenly tuning the coupler close to the avoided crossing induces non-adiabatic transitions, leading to population exchange between interacting states. These transitions introduce additional amplitude and phase oscillations during the interaction on timescales of nanoseconds. 
 We employ a Savitzky–Golay filter with a window length of \SI{10}{\nano\second} to smooth out these oscillations, thereby limiting the resolution of the flux response measurement.
To calibrate the applied flux amplitude to the resulting qubit frequency change, we measure the frequency shift induced by various nominal flux amplitudes for pulse lengths $>\SI{300}{\nano\second}$.
The non-linear relation between the applied flux and the qubit frequency  accurately resolves the coupler's response near the qubit, while suppressing the contribution of turn-off transients when the coupler is detuned from the qubit.

\begin{table}[tb]
    \centering
    \begin{tabular}{l c c c c}
        $i$      & 1         & 2         & 3         & 4         \\ \hline
            $A_i$    & $-0.021$  & $-0.012$  & $-0.393$  & $+0.595$  \\ \hline
            $\tau_i$ (ns) & $846$     & $151$     & $36.0$    & $21.6$ 
    \end{tabular}
    \caption{
        Fitted parameters for the exponential IIR filter corrections.\label{tab:exp_filter_coeff}
    }
\end{table}

The measured flux-pulse step response  (blue curve in Fig. \ref{fig:Distortion}(b)) exhibits low-pass filtering characteristics, influenced by various factors. These include the skin effect in the cabling~\cite{wigington1957transient,nahman1962discussion}, which induces distortions on longer time scales, as well as filters, packaging, and on-chip distortions that predominantly impact shorter timescales.
After \SI{60}{\nano\second}, the step response reaches \SI{97}{\percent} of the final flux value while  reaching \SI{99}{\percent} requires over \SI{650}{\nano\second}, with residual deviations persisting for up to \SI{2.5}{\micro\second}.

The measured flux response is modeled using  exponential IIR filters of the form
\begin{align}
    y(t) = y_\text{in}(t) \cdot \prod_i(1 + A_i e^{-t/\tau_i}),
\end{align}
where $A_i$ are the amplitudes and $\tau_i$ are the decay constants.
Long-time-scale distortions are corrected using two exponential IIR filters ($i\in [1,2]$), resulting in the red curve in Fig.~\ref{fig:Distortion}(b).
Based on this corrected signal, the short-time-scale behavior is addressed using two additional exponential IIR filters ($i\in{3,4}$).
 The parameters of the exponential filters are provided in Table~\ref{tab:exp_filter_coeff}.
These FPGA-based real-time distortion compensations of the HDAWGs~\cite{Zurichinstrumentsag2022}, are constrained to a granularity of \SI{4}{\nano\second} by the FPGA clock rate.
The initial samples are further corrected using an FIR filter, which convolves the signal as follows
\begin{align}
    y[n]=\sum_{m=0}^{M-1} l[m] \cdot y_\text{in}[n-m],
\end{align}
with $M=72$ calibrated coefficients $l[m]$.
This process results in the fully corrected signal, shown as the orange curve in Fig. \ref{fig:Distortion}(b), achieved using the full set of four IIR filters ($i\in {1,2,3,4}$) and the FIR filter. The corrected response aligns with the target response function within $\pm 0.5\,\%$, with no resolvable distortions after $\SI{60}{\nano\second}$.

\subsection{Gate fidelity effects}\label{app:flux_dist_effects}
\begin{figure}[tb]
    \centering
    \includegraphics[]{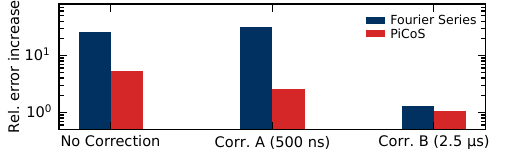}
    \caption{
    Relative increase in gate errors when using $n=2$ CZ gates in comparison to $n=1$ CZ gates in interleaved randomized benchmarking, seperated for Fourier Series and PiCoS pulses.
    The following cases with no correction, and flux distortion corrections measured and corrected for up to (A) $\SI{500}{\nano\second}$, and (B) $\SI{2.5}{\micro\second}$.  
    }
    \label{fig:RelativeSecondaryError}
\end{figure}
Since flux pulses are exclusively used within CZ gates, we evaluate the impact of flux distortions by interleaving two CZ gates -- instead of a single CZ gate -- between the Clifford gates in the interleaved randomized benchmarking sequence, such that the flux pulses directly follow each other.
With no correction the total errors of the multi-gate measurement are \SI{16.6\pm 2.4}{\percent} (up from the individual error \SI{0.62\pm 0.15}{\percent}) for the Fourier-series pulses, respectively. The relative increase in errors is shown in Fig.~\ref{fig:RelativeSecondaryError}.
In contrast, the PiCoS pulse exhibits only a moderately increased error of \SI{1.6\pm0.4}{\percent} (up from the individual error \SI{0.52\pm 0.19}{\percent}).
This indicates that the PiCoS parameterization effectively reduces the impact of signal distortions. However, it cannot mitigate distortion effects that occur outside the pulse duration.

 With the additional on-device flux distortion correction, 
 the errors of the optimized pulses are reduced, particularly mitigating the error increase observed after the second gate.
 Our analysis demonstrates the necessity of integrating the complete system response into the evaluation. 
 During our experiments, the flux pulse response was measured and corrected twice: first with a \SI{500}{\nano\second} flux pulse length, and later with a \SI{2.5}{\micro\second} flux pulse length.
The initial correction reduced flux distortions, leading to high-fidelity CZ gates, assessed through interleaved randomized benchmarking. However, errors accumulated when probing two successive interleaved CZ gates, suggesting immediately adjacent flux pulses still degrate the average gate performance, and corrections on longer timescales are needed (see Corr. A. in Fig.~\ref{fig:RelativeSecondaryError}).
Later, the response function was measured over a longer time window of \SI{2.5}{\micro\second}, close to the maximum time that could be reliably resolved in the presence of decoherence and population transfer into coupler states. This correction effectively eliminated the error increase for successive CZ gates, as shown by Corr. B in  Fig.~\ref{fig:RelativeSecondaryError}.
The results also show that the PiCoS pulse minimizes additional errors across all cases. 
 
The error due to flux-pulse bleeding is most pronounced in the accumulated phase $\varphi_1$ of $Q_1$, which is highly sensitive to the exact coupler-frequency trajectory during the pulse. Even minor changes caused by pulse bleeding can result in significant variations in the acquired phases due to the strong dependence of the state frequency on the coupler frequency, as shown in  Fig.~\ref{fig:system_chip}(b).

Note that the immediate concatenation of two CZ gates on the same control line is rarely encountered in practical applications. The entangling gates are typically alternated with single-qubit operations or operations on other qubit pairs, which naturally mitigates long-term effects~\cite{Sack2024,Tilly2022}.

\end{document}